\documentclass[letterpaper,twocolumn,10pt]{article}
\usepackage{usenix2019_v3}
\usepackage{amssymb}
\usepackage{pifont}
\usepackage{graphicx}  
\usepackage{amsmath} 
\usepackage{cuted} %
\usepackage{paralist}  
\usepackage{enumitem}  

\usepackage{booktabs}
\usepackage{makecell}
\usepackage[table]{xcolor}
\usepackage{pifont}

\newcommand{\cmark}{\ding{51}}
\newcommand{\xmark}{\ding{55}}
\rowcolors*{2}{blue!3}{white}  %

\usepackage{bbding}
\usepackage{amsmath}
\usepackage{amssymb}
\usepackage{amsfonts}
\usepackage{algorithm}
\usepackage{algpseudocode}
\usepackage{footnote}
\usepackage{booktabs}
\usepackage{siunitx}  
\usepackage{geometry}
\usepackage{natbib}

\usepackage{fancybox}
\usepackage{tcolorbox}

\usepackage{tikz}
\usetikzlibrary{shapes, arrows, positioning}
\usepackage{comment}
\usepackage{amsthm}
\usepackage{subcaption}
\usepackage{multirow}
\usepackage{comment}
\usepackage{makecell} 
\usepackage{tabularx}
\usepackage[switch]{lineno}

\usepackage{xspace}
\usepackage[export]{adjustbox}
\usepackage{graphicx}
\usepackage{svg}
\usepackage{enumitem}
\setlist[itemize]{itemsep=0pt, topsep=0pt}

\newcommand{\eat}[1]{}

\newcommand{\cbit}{\begin{compactitem}}
\newcommand{\ceit}{\end{compactitem}}
\newcommand{\cben}{\begin{compactenum}}
\newcommand{\ceen}{\end{compactenum}}

\newcommand{\logofast}{\includegraphics[width=0.02\linewidth]{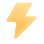}}
\newcommand{\logoslow}{\includegraphics[width=0.02\linewidth]{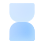}}

\usepackage[capitalize]{cleveref}
\crefname{section}{Sec.}{Secs.}
\crefname{table}{Tab.}{Tables}
\crefname{figure}{Fig.}{Figs.}
\crefname{algocf}{alg.}{algs.}
\Crefname{algocf}{Algorithm}{Algorithms}

\begin{document}

\date{}

\title{\Large \bf A.S.E: A Repository-Level Benchmark for Evaluating Security in AI-Generated Code}

\author{%
\begin{minipage}{\textwidth}\centering
{\rm
Keke Lian$^{1\dagger}$, %
Bing Wang$^{2\dagger}$, %
Lei Zhang$^{3}$, %
Libo Chen$^{4}$, %
Junjie Wang$^{5}$, %
Ziming Zhao$^{6}$,  
Yujiu Yang$^{5}$, %
Miaoqian Lin$^{1}$, %
Haotong Duan$^{1}$, %
Haoran Zhao$^{3}$, %
Shuang Liao$^{3}$, %
Mingda Guo$^{3}$, %
Jiazheng Quan$^{2}$, 
Yilu Zhong$^{2}$, %
Chenhao He$^{4}$, %
Zichuan Chen$^{4}$, %
Jie Wu$^{5}$, %
Haoling Li$^{5}$, %
Zhaoxuan Li$^{7}$, 
Jiongchi Yu$^{8}$, %
Hui Li$^{2*}$, %
Dong Zhang$^{1*}$%
}\\[0.6ex]
{\small
$^{1}$Tencent \quad
$^{2}$Peking University \quad
$^{3}$Fudan University \quad
$^{4}$Shanghai Jiao Tong University \\ 
$^{5}$Tsinghua University \quad
$^{6}$Zhejiang University \quad
$^{7}$Institute of Information Engineering, Chinese Academy of Sciences \quad
$^{8}$Singapore Management University
}
\end{minipage}
}

\maketitle

\begin{strip}
\centering
\small
\textbf{A.S.E Code:} \url{https://github.com/Tencent/AICGSecEval}
 \\
\textbf{A.S.E  Website:} \url{https://aicgseceval.tencent.com/home}
\end{strip}

\begingroup
\renewcommand\thefootnote{\fnsymbol{footnote}}
\footnotetext[2]{Co-first authors.}
\footnotetext[1]{Corresponding authors: Hui Li (\texttt{lih64@pkusz.edu.cn}) and Dong Zhang (\texttt{zalezhang@tencent.com}).}
\endgroup

\begin{abstract}
The increasing adoption of large language models (LLMs) in software engineering necessitates rigorous security evaluation of their generated code. However, existing benchmarks often lack relevance to real-world AI-assisted programming scenarios, making them inadequate for assessing the practical security risks associated with AI-generated code in production environments. To address this gap, we introduce A.S.E (AI Code Generation Security Evaluation), a repository-level evaluation benchmark designed to closely mirror real-world AI programming tasks, offering a comprehensive and reliable framework for assessing the security of AI-generated code. Our evaluation of leading LLMs on A.S.E reveals several key findings. In particular, current LLMs still struggle with secure coding. The complexity in repository-level scenarios presents challenges for LLMs that typically perform well on snippet-level tasks. Moreover, a larger reasoning budget does not necessarily lead to better code generation. These observations offer valuable insights into the current state of AI code generation and help developers identify the most suitable models for practical tasks. They also lay the groundwork for refining LLMs to generate secure and efficient code in real-world applications.
\end{abstract}

\section{Introduction}
The rapid advancement of large language models (LLMs) has greatly enhanced the AI programming ecosystem, with tools like Cursor~\cite{@Cursor} and Claude Code~\cite{@claude-code} enabling developers to choose models that best fit their tasks. These AI assistants significantly improve programming efficiency, leading to a surge of AI-generated code in production environments. However, research~\cite{@8SecurityEval,DBLP:journals/corr/abs-2502-11844@12BaxBench,DBLP:conf/llm4code/PengCHYR25@11CWEval,DBLP:conf/satml/HajipourHHSF24@9CodeLMSec,DBLP:journals/corr/abs-2506-05692@13SafeGenBench,fu2024constraineddecodingsecurecode, he2023large, pearce2025asleep, wang2024your} has shown that such code can harbor security vulnerabilities, posing serious risks such as data breaches or system failures~\cite{pearce2025asleep, @8SecurityEval, fu2023security, li2017large}. Relying on developers to ensure the security of AI-generated code can be highly challenging given the complexity of modern software systems. Therefore, \textit{there is a pressing need to identify and utilize AI models that are capable of generating secure code}.

Despite the numerous benchmarks~\cite{DBLP:conf/nips/HendrycksBKMAGB21@21APPS, dou2024stepcoder, DBLP:journals/corr/abs-2107-03374@17HumanEval, DBLP:journals/corr/abs-2108-07732@19MBPP, DBLP:journals/corr/abs-2506-05692@13SafeGenBench, DBLP:journals/corr/abs-2502-11844@12BaxBench, DBLP:conf/llm4code/PengCHYR25@11CWEval, DBLP:journals/corr/abs-2405-00218@10constrained, DBLP:conf/satml/HajipourHHSF24@9CodeLMSec, @8SecurityEval, wang2024your} developed by both academia and industry to evaluate AI-generated code, most of them ~\cite{DBLP:conf/nips/HendrycksBKMAGB21@21APPS, dou2024stepcoder, DBLP:journals/corr/abs-2107-03374@17HumanEval, DBLP:journals/corr/abs-2108-07732@19MBPP} primarily focus on code quality, such as syntax correctness and functional accuracy, while overlooking critical security considerations. 
Although some benchmarks~\cite{DBLP:journals/corr/abs-2506-05692@13SafeGenBench, DBLP:journals/corr/abs-2502-11844@12BaxBench, DBLP:conf/llm4code/PengCHYR25@11CWEval, DBLP:journals/corr/abs-2405-00218@10constrained, DBLP:conf/satml/HajipourHHSF24@9CodeLMSec, @8SecurityEval, wang2024your} attempt to address code security (as shown in \autoref{tab:security_benchmarks}), they are often \textit{inadequate} to assess the actual security risks of AI-generated code in real-world production scenarios due to several key reasons:
\textbf{(a) Limited relevance to real-world data.} These datasets are typically sourced from human-curated synthetic code snippets, which have limited relevance to the functional and security scenarios of real-world projects.
\textbf{(b) Code generation tasks detached from real-world AI programming.} Their code generation tasks are typically limited to isolated code snippets, focusing solely on functional descriptions without considering the context within files or projects, which does not align with mainstream AI programming paradigm.
\textbf{(c) Unreliable code evaluation methods.} Security assessments of generated code often rely on manual or LLM-based judgment, which are unreliable and difficult to automate or reproduce consistently.
This gap poses a significant challenge for both developers and organizations seeking to integrate AI-generated code securely into their systems.

To bridge this gap, we introduce A.S.E (AI Code Generation Security Evaluation), a repository-level evaluation benchmark designed to closely mirror real-world AI programming scenarios, offering a comprehensive and reliable framework for assessing the security of AI-generated code.
Specifically, A.S.E has the following key design features: \textbf{(a) Real-world data source}: The dataset is derived from high-quality GitHub open-source repositories with documented CVEs. A.S.E leverages vulnerability-related code extracted from CVE patches, ensuring that the data reflects both realistic and security-sensitive scenarios. 
\textbf{(b) Simulation of real-world code generation tasks}: A.S.E mimics AI programming assistants like Cursor by extracting code contexts (including both intra-file and cross-file contexts) from the repository, and providing them to LLMs for code generation. 
\textbf{(c) High accuracy and reproducibility in code evaluation}: For each test case, corresponding to a specific CVE and repository, A.S.E designs targeted static vulnerability detection rules that can scan for the original CVE, ensuring accurate security assessment of the regenerated project.

Building on these design principles, the A.S.E benchmark includes 120 repository-level instances, consisting of 40 seed dataset collected from GitHub and 80 mutated variants generated through semantic and structural mutation techniques, such as identifier renaming and control-flow reshaping. These variants are introduced to mitigate data leakage risks, ensuring that the evaluation reflects the LLM's capabilities rather than its memorization. 
To closely simulate real-world code generation processes, A.S.E uses the BM25 algorithm~\cite{DBLP:journals/ftir/RobertsonZ09@bm25} to automatically extract relevant code context from the repository when designing code generation tasks. These code generation tasks cover four common vulnerability types in real-world web projects. We focus on the web domain due to the primary application of AI programming in web development in production environments. These vulnerability-related codes to be generated are closely tied to the projects' business logic, requiring models to understand both project code and security considerations. A.S.E also incorporates five commonly used web development languages to assess how programming languages influence model performance. 
Furthermore, A.S.E integrates customized static application security testing (SAST) rules for security evaluation on each data instance in the dataset. While A.S.E aims to evaluate the security of AI-generated code, it also considers code quality and stability. Given that security is meaningful only when code correctness is ensured, and in light of potential hallucinations in large model outputs, A.S.E uses multi-dimensional metrics that assess code security, quality, and generation stability for a comprehensive evaluation of the model’s capabilities.

Based on A.S.E, we evaluated 26 mainstream commercial or open-source models under the same experimental setup, leading to several key findings. First, existing LLMs still face significant challenges in secure coding. All models fall short in terms of security performance compared to their code quality performance (such as syntax correctness). Even the best-performing model, Claude-3.7-Sonnet, achieved only a total score of 52.79 in our evaluation. Second, A.S.E introduces significant complexity in repository-level scenarios, which presents a challenge for LLMs that typically perform well on snippet-level tasks. For example, although GPT-3 excels on SafeGenBench~\cite{DBLP:journals/corr/abs-2506-05692@13SafeGenBench}, its performance on the A.S.E. benchmark drops, falling behind many other models. Third, slow-thinking configurations, which allocate more deliberate computation or multi-step reflection, tend to underperform compared to fast-thinking configurations that rely on concise, direct decoding. This suggests that a larger reasoning budget does not necessarily lead to better code generation.
These observations offer valuable insights into the current state of AI code generation, helping developers select the most appropriate models for their specific tasks. Furthermore, they provide a foundation for refining LLMs, enhancing their ability to generate secure and efficient code in real-world applications.

The main contributions of this paper are summarized as follows:
\begin{itemize}[leftmargin=1.2em]
\item \textbf{New repository-level benchmark from real code.} 
We release A.S.E, a repository-level evaluation benchmark derived from real-world GitHub repositories with documented CVEs. Unlike existing benchmarks, A.S.E is designed to closely mirror real-world AI programming tasks by leveraging vulnerability-related code from CVE patches, ensuring the data reflects both realistic and security-sensitive scenarios.
\item \textbf{Automated and reproducible evaluation framework.} 
We develop a reproducible vulnerability-targeted evaluation framework that integrates custom vulnerability detection rules tailored to each data instance. Compared to previous work, this framework enables more automated and accurate code evaluation. It comprehensively considers the capabilities of AI-generated code, including security, quality, and generation stability.
\item \textbf{Extensive experiments and findings.} 
We evaluate 26 mainstream commercial and open-source LLMs on A.S.E, revealing several key findings. These insights shed light on the current state of AI code generation, guiding developers in selecting the most suitable models for their tasks. Additionally, they lay the groundwork for refining LLMs to improve their ability to generate secure and efficient code in real-world applications.
\end{itemize}

\section{Related Work}

In recent years, numerous benchmarks have been developed for evaluating AI-generated code, which can be broadly categorized into two types. Functionality-oriented benchmarks~\cite{DBLP:journals/corr/abs-2107-03374@17HumanEval,DBLP:journals/corr/abs-2108-07732@19MBPP,DBLP:conf/iclr/0003XM24@28RepoBench,DBLP:journals/corr/abs-2406-11612@31Long-code-arena,DBLP:conf/nips/DingWADTJRNBRX23@29CrossCodeEval,DBLP:conf/acl/LiangJH025@32REPOCOD,DBLP:conf/acl/LiZGM0PHWL25@33FEA-Bench,jimenez2024swebench}--the majority of existing efforts--primarily assess whether the generated code is syntactically correct and functionally accurate, with limited emphasis on security or vulnerability detection. For example, HumanEval~\cite{DBLP:journals/corr/abs-2107-03374@17HumanEval} evaluates models through algorithmic-contest programming tasks to measure functional correctness in language understanding, algorithms, and basic mathematics. Security-oriented benchmarks~\cite{@8SecurityEval,DBLP:journals/corr/abs-2502-11844@12BaxBench,DBLP:conf/llm4code/PengCHYR25@11CWEval,DBLP:conf/satml/HajipourHHSF24@9CodeLMSec,DBLP:journals/corr/abs-2506-05692@13SafeGenBench,fu2024constraineddecodingsecurecode}, on the other hand, go a step further by evaluating whether the generated code is secure and reliable. \autoref{tab:security_benchmarks} summarizes representative benchmarks in this category and compares them with A.S.E. 
In what follows, we highlight the distinctions between A.S.E and existing benchmarks from two perspectives: relevance to real-world scenarios and code assessment methods.

\begin{table*}[!t]
\centering
\caption{Comparison of Security-Oriented Code Generation Evaluation Datasets.}
\label{tab:security_benchmarks}
\renewcommand{\arraystretch}{1.25}

\rowcolors{2}{blue!3}{white}

\resizebox{\textwidth}{!}{
\begin{tabular}{lcccccc}
\toprule
\makecell{\textbf{Dataset}} &
\makecell{\textbf{CWE}\\\textbf{Tags}} &
\makecell{\textbf{Granularity}\\\textbf{(Repo/Snippet)}} &
\makecell{\textbf{Provenance}} &
\makecell{\textbf{Domain}} &
\makecell{\textbf{Open}\\\textbf{Source}} &
\makecell{\textbf{Security}\\\textbf{Eval.}} \\
\midrule
\rowcolor{blue!15}
A.S.E (Ours)       & \cmark & \textbf{Repository}    & Real-World Repos              & \makecell{Realistic Full-Web Repositories} & \cmark & SAST \\
SafeGenBench       & \cmark & Snippet & Human-Curated Synthetic & \makecell{Simplified Programming Tasks}      & \cmark & SAST + LLM \\
BaxBench           & \xmark & Snippet & Human-Curated Synthetic & \makecell{Backend Programming Tasks}       & ---    & Test Cases \\
CWEval             & \cmark & Snippet & Human-Curated Synthetic & \makecell{Simplified Programming Tasks}         & \cmark & Test Cases \\
CODEGUARD+         & \cmark & Snippet & Human-Curated Synthetic & \makecell{Simplified Programming Tasks}      & ---    & Test Cases \\
CodeLMSec          & \cmark & Snippet & Human-Curated Synthetic & \makecell{Simplified Programming Tasks}      & \cmark & SAST \\
SecurityEval       & \cmark & Snippet & Human-Curated Synthetic & \makecell{Simplified Programming Tasks}      & \cmark & SAST + Manual \\
\bottomrule
\end{tabular}
}
\rowcolors{0}{}{}
\end{table*}

\subsection{Relevance to Real-world Scenarios}

Early functionality-oriented benchmarks for code generation typically emphasized small, self-contained tasks at the function or snippet level (e.g., HumanEval~\citep{DBLP:journals/corr/abs-2107-03374@17HumanEval}, MBPP~\citep{DBLP:journals/corr/abs-2108-07732@19MBPP}), using unit tests or reference outputs to measure basic functional correctness. Their datasets are often constructed from human-curated synthetic sources, such as algorithm competition problems, which provide well-defined but limited scenarios. Due to their simplicity, standard code benchmarks have quickly become saturated. Consequently, more challenging benchmarks with stronger relevance to real-world scenarios have emerged~\cite{DBLP:conf/iclr/0003XM24@28RepoBench,DBLP:journals/corr/abs-2406-11612@31Long-code-arena,DBLP:conf/nips/DingWADTJRNBRX23@29CrossCodeEval,DBLP:conf/acl/LiangJH025@32REPOCOD,DBLP:conf/acl/LiZGM0PHWL25@33FEA-Bench,jimenez2024swebench}. For example, SWE-Bench~\cite{jimenez2024swebench} evaluates the code generation capabilities of LLMs and agents by framing tasks around issue resolution in real-world projects.

In contrast, security-oriented benchmarks remain at a relatively early stage. Their scope is still limited, with most focusing on snippet-level tasks that lack strong connections to real-world programming scenarios~\citep{@8SecurityEval,DBLP:journals/corr/abs-2502-11844@12BaxBench,DBLP:conf/llm4code/PengCHYR25@11CWEval,DBLP:conf/satml/HajipourHHSF24@9CodeLMSec,DBLP:journals/corr/abs-2506-05692@13SafeGenBench,fu2024constraineddecodingsecurecode}. This restricts their ability to capture the complexity of security issues in practical development environments. 
To address this gap, we introduce A.S.E, which, similar to SWE-Bench, is derived from real-world projects. Unlike existing security-oriented benchmarks, A.S.E builds on repositories with documented CVEs and incorporates repository-level context into the code generation process, thereby ensuring that evaluations are both realistic and security-sensitive.

\subsection{Code Assessment Methods}
Prior security-oriented benchmarks have adopted diverse methods for assessing the security of AI-generated code, yet each comes with notable limitations. 
(a) Manual evaluation offers flexibility but is labor-intensive and difficult to scale. For example, SecurityEval~\cite{@8SecurityEval} relies on expert validation, which makes it difficult to automate or scale for evaluating new models. 
(b) LLM-as-judge approaches scale more easily and capture semantic nuances, but their judgments are highly sensitive to prompt design, model version, and decoding randomness, making results unreliable and hard to reproduce~\citep{DBLP:journals/corr/abs-2312-04724@25CyberSecEval}. 
(c) Generic static analysis (SAST) tools provide deterministic rule-based detection, yet often suffer from false positives and false negatives when applied across different languages or without contextual calibration~\cite{DBLP:conf/satml/HajipourHHSF24@9CodeLMSec,DBLP:journals/corr/abs-2506-05692@13SafeGenBench}.
(d) Test-case-based assessments are straightforward but inherently limited~\cite{DBLP:journals/corr/abs-2502-11844@12BaxBench,DBLP:conf/llm4code/PengCHYR25@11CWEval,fu2024constraineddecodingsecurecode}. For instance, BaxBench~\cite{DBLP:journals/corr/abs-2502-11844@12BaxBench} designs correctness and security tests that are agnostic to frameworks and programming languages, but this generality makes it difficult to fully validate generated code, leading to under-reporting of vulnerabilities.
To overcome these challenges, A.S.E emphasizes reproducibility and precision through customized vulnerability detection. For each data instance corresponding to a specific project and CVE, A.S.E designs tailored static analysis rules that explicitly model the relevant sources, sinks, and taint propagation paths. These rules are calibrated to accurately detect the original vulnerability within the unmodified repository, ensuring that the evaluation framework can reliably assess whether regenerated code reintroduces or mitigates the same flaw. By automating this process at the project level, A.S.E achieves more reliable, scalable, and reproducible security assessments compared to prior approaches.


\section{The A.S.E Benchmark}

\begin{figure*}[!tp]
\centering
\includegraphics[width=\linewidth]{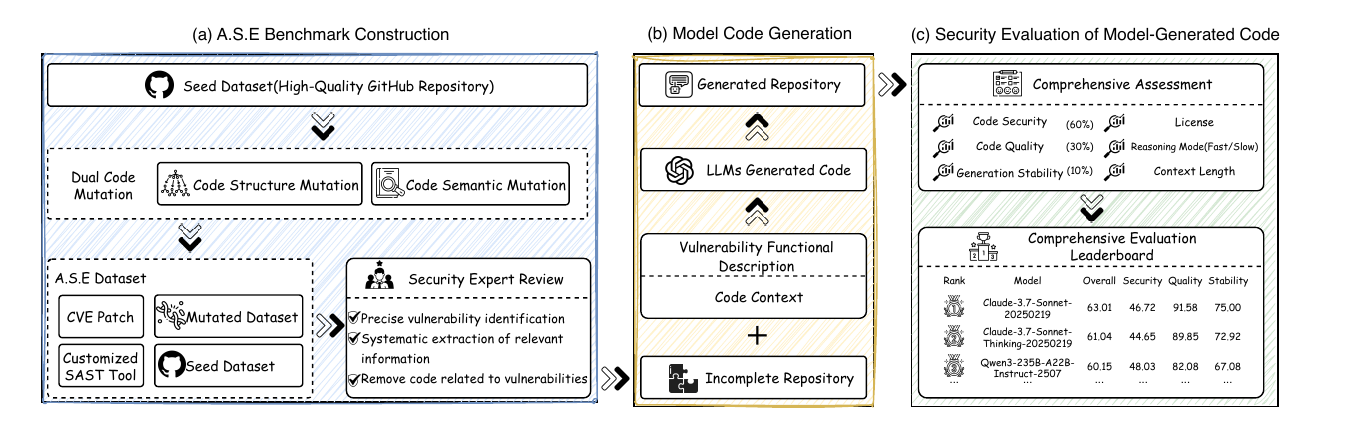} 

\caption{Overall workflow of A.S.E. (a) A.S.E benchmark construction: from high-quality GitHub seeds, we build the A.S.E dataset via dual mutations (structure/semantic), CVE patches, and a customized SAST tool, followed by expert curation. (b) Model code generation: given an incomplete repository, a vulnerability description and context guide LLMs to complete the repository. (c) Security evaluation: comprehensive assessment with security, quality and stability.}
\label{fig:ase_workflow}
\end{figure*}

This section introduces the A.S.E. framework, which includes three core components: benchmark construction (\autoref{ss:benchmark-construction}), code generation task setup (\autoref{ss:codegenerationtask}), and code evaluation (\autoref{ss:codeassessment}), as shown in \autoref{fig:ase_workflow}.
We will first highlight the key features of the A.S.E. design, followed by a detailed discussion of each of these three core components.

\subsection{Design Philosophy}
\label{ss:design-philosophy}

We aim to create a repository-level evaluation benchmark that mirrors real-world AI programming scenarios, offering a reliable framework for assessing the security of AI-generated code. To ensure accurate results, A.S.E. focuses on realistic data sources, task settings, and code assessment methods.
Specifically, the core features of the A.S.E benchmark are designed around the following principles:

\textbf{(i) Data Source: Real-world and Repository-Level Data Sources.}
To reflect the performance of large models in real-world software environments, A.S.E constructs tasks from active open-source repositories with documented CVEs and verifiable patches. It utilizes vulnerability-related code extracted from CVE patches, ensuring the data captures realistic and security-sensitive scenarios. To mitigate the risk of data leakage, A.S.E employs semantic and structural mutation techniques, such as identifier renaming and control-flow reshaping, on the collected real-world repositories. These variants help prevent data leakage and ensure that the evaluation reflects the LLM's capabilities, not its memorization.

\textbf{(ii) Task Settings: Practical Simulations of AI Programming Workflows.}
To simulate realistic usage, A.S.E replicates AI programming assistants like Cursor by extracting code contexts—both intra-file and cross-file—directly from the repository. These contexts are then provided to LLMs for code generation, closely mimicking real-world AI programming scenarios. Moreover, the generated code is output in the form of \texttt{diff} files, allowing patches to be applied directly to the repository, further reflecting real software development practices.

\textbf{(iii) Code Assessment: High Accuracy and Reproducibility Assessment.}
Instead of relying on manual or LLM-based judgment, which can be unreliable and difficult to reproduce consistently, A.S.E designs targeted static vulnerability detection rules for each test case, corresponding to a specific CVE and repository. These rules are tailored with dedicated source–sink definitions and taint propagation paths to successfully detect the original CVE, thereby ensuring an accurate security assessment of the regenerated project.

Following these guidelines, we introduce the A.S.E benchmark to 
evaluate in repository-level code generation regarding security. 
After that, we detail the three key steps: benchmark construction, task design, and result evaluation.

\begin{figure*}[!tp]
\centering
\includegraphics[width=\linewidth]{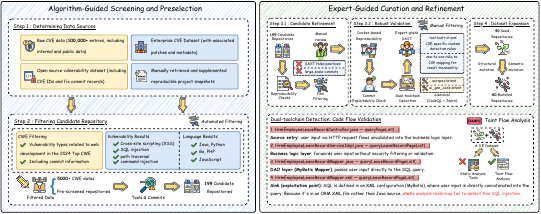}  
\caption{Overview of A.S.E benchmark construction. \textbf{Algorithm-guided screening and preselection (left):} aggregate CVE-linked sources and automatically filter repositories by web-related CWEs, vulnerability types (XSS, SQL injection, path traversal, command injection), and languages (Java, Python, Go, PHP, JavaScript). \textbf{Expert-guided curation and refinement (right):} conduct manual review, reproducibility and exploitability checks, and dual-toolchain SAST (e.g., CodeQL + Joern) with CVE-specific rules; then expand $40$ seed repositories via structural/semantic mutation to $80$ variants. }
\label{fig:benchmark_construction}
\end{figure*}

\subsection{A.S.E Benchmark Construction}
\label{ss:benchmark-construction}
Guided by our design philosophy, we construct the A.S.E benchmark as shown in~\autoref{fig:ase_workflow} (a) and~\autoref{fig:benchmark_construction}.
To ensure realistic scenarios and relevant security expertise, we organize a dedicated team of ten contributors with strong backgrounds in cybersecurity and web development from top-tier universities, including five Ph.D. candidates and five master-level students. All contributors have extensive hands-on experience in vulnerability discovery, analysis, and remediation, with emphasis on common web vulnerabilities such as XSS, SQL injection, and path traversal. Each contributor adheres to strict secure coding standards and is familiar with static code analysis techniques.
The construction of the benchmark proceeds in four stages: determining data sources, filtering candidate repository, expert-guided refinement and quality filtering, and dataset expansion.
The following subsections will provide a detailed explanation of each step.

\textbf{Step 1: Determining Data Sources.}
To ensure that our code generation tasks are both security-sensitive and representative of real-world scenarios, we begin by collecting source data directly from CVE vulnerabilities. Specifically, we gather CVE records and their corresponding repository information from both public vulnerability databases and enterprise-internal repositories. For each repository, we additionally require accessible commit history, which allows us to accurately locate the vulnerable code and later design code generation tasks.
As a result of this initial step, we obtain more than $100,000$ raw CVE entries as the starting point for benchmark construction.

\textbf{Step 2: Filtering Candidate Repositories.} 
After collecting raw CVE entries, we perform a strict filtering process to ensure both project quality and vulnerability completeness. 
Specifically, we first narrow the dataset by retaining only entries that (i) map to web-relevant categories in the 2024 Top CWE list~\cite{CWE_Top25_2024} and (ii) provide traceable vulnerability fix commit contexts. This restriction reflects our focus on web applications, which constitute one of the most common real-world domains for AI programming, and ensures that every vulnerability can be grounded in a concrete, analyzable code change rather than an abstract description. Then we further filter the repositories by requiring either active monthly maintenance or a popularity threshold of more than $1,000$ GitHub stars. This step eliminates abandoned or low-quality projects and ensures that the retained repositories exhibit sufficient code complexity and practical relevance, thereby making the benchmark more representative of real-world software engineering environments. After applying these criteria, the dataset is reduced to approximately $50,000$ candidate repositories.

Furthermore, we apply SAST tools (e.g., CodeQL~\cite{CodeQL} and Joern~\cite{Yamaguchi2014ModelingAD}) to these repositories and intersect reported findings with the lines modified in the corresponding commits. By retaining only cases where SAST detections coincide with the lines modified in the corresponding commits, we simultaneously reduce false positives and enforce a verifiable vulnerability–fix causal chain. This design ensures that each CVE instance is (i) detectable by automated static analysis, confirming its practical observability, and (ii) accurately linked to the correct fixing commit, guaranteeing that the patched code genuinely corresponds to the target vulnerability. These conditions provide reliable vulnerability evidence that can support subsequent expert-guided analysis. As a result, this stage yields 199 high-confidence candidates.

\textbf{Step 3: Expert-Guided Refinement and Quality Control.} 
To ensure the final benchmark is genuinely security-relevant and reflects real-world vulnerabilities that are both detectable and reproducible, we refine the dataset through expert annotation and validation.
Specifically, we first conduct an initial manual review to further control data quality. At this stage, security experts remove obvious false positives introduced by static analysis tools and discard commits that modify an excessive number of files (e.g., more than 10), since large-scale changes obscure vulnerability localization and hinder precise labeling. Building on the cleaned candidate set, we then proceed with a fine-grained expert analysis that focuses on the vulnerabilities themselves. Security experts precisely annotate the vulnerable code regions, reconstruct the relevant execution context (e.g., source/sink signatures, API definitions, call chains), and design targeted CodeQL/Joern queries to validate taint propagation paths. 
Once validated, the labeled vulnerable code is removed to create a fill-in-the-code setting. By combining the functional description of the vulnerability with its extracted context, we generate structured prompts that require models to reason over repository-level structures and logic rather than isolated snippets. After final expert review, 40 repositories with verified CVE records are retained as the seed dataset, each anchored at a baseline commit that provides a stable starting point for task construction and evaluation. This expert-driven process guarantees authenticity, reliability, and reproducibility across all benchmark tasks.

\textbf{Step 4: Dataset Expansion.}
After constructing base tasks from real-world CVE vulnerabilities and corresponding repositories, we expand the dataset to increase both volume and coverage under a semantics-preserving constraint. Two categories of transformations are applied. The first is semantic transformations, which diversify surface expressions by systematically renaming variables and functions or substituting equivalent APIs. The second is structural transformations, which alter control flow, refactor call graphs, or reorganize file layouts to introduce structural variation. These operations preserve functional behavior and code semantics while modifying implementation details, thereby reducing overlap with public repository code that may appear in model training corpora and mitigating potential data contamination. The expanded variants enable a more comprehensive assessment of model robustness and generalization. 
Through these transformations, we generate 80 additional variants from the 40 seed repositories, yielding a total of 120 benchmark instances.

\begin{figure*}[!tp]
\centering
\includegraphics[width=1\linewidth]{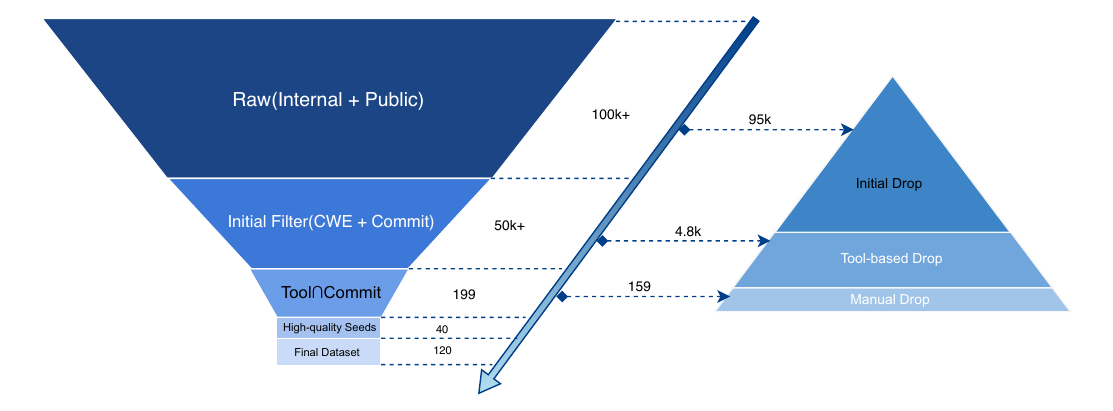}
\caption{Benchmark Construction Funnel. }
\label{fig:ASE.benchmark-loudou}
\end{figure*} 

\begin{figure*}[!tp]
\centering
\includegraphics[width=\linewidth]{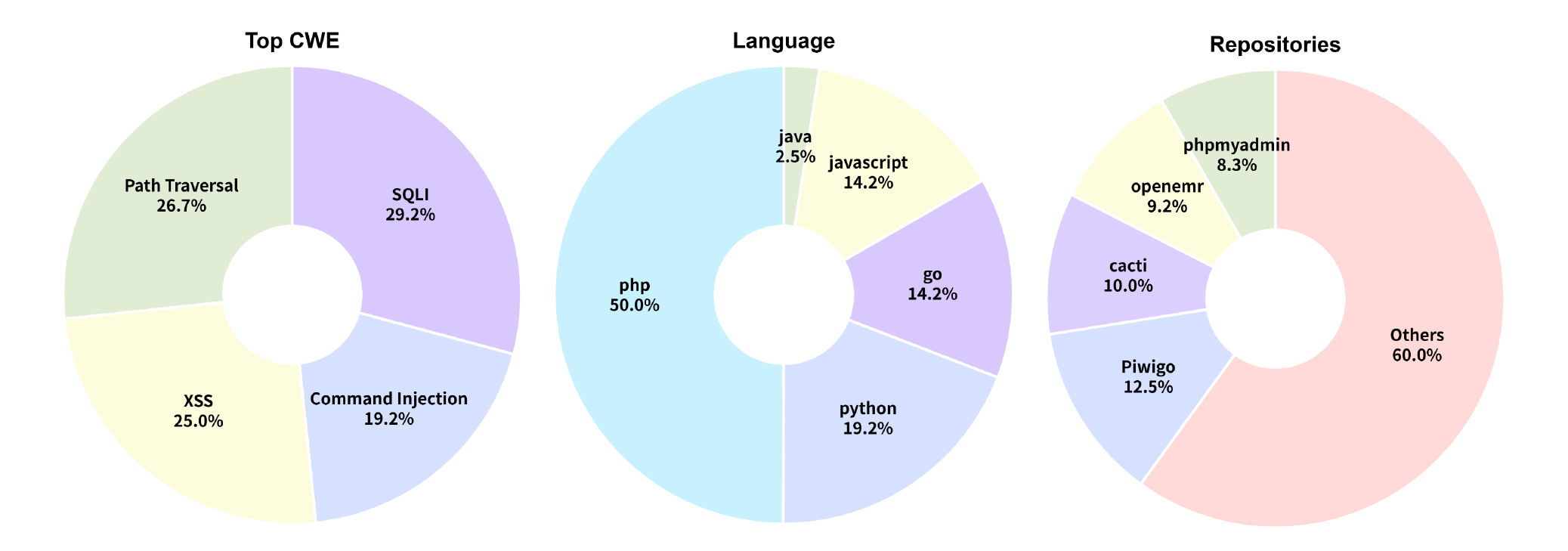}
\caption{Statistics of A.S.E benchmark, including the distribution of top CWE categories, programming languages, and repositories.}
\label{fig:ASE.benchmark-pie}
\end{figure*}

\vspace{5pt}
\noindent
\textbf{General Statistics.}
\autoref{fig:ASE.benchmark-loudou} illustrates the data reduction pipeline from the initial collection of raw CVE entries to the final benchmark. The funnel chart presents the number of instances retained after each stage of filtering and refinement, showing how the dataset was progressively narrowed from a large pool of raw vulnerabilities to a carefully curated benchmark.
The resulting A.S.E benchmark comprises $120$ repository-level vulnerability instances and the overall composition is illustrated in \autoref{fig:ASE.benchmark-pie}. 

Specifically, the dataset targets four categories of vulnerabilities that are widely prevalent in real-world web projects, each aligned with a CWE entry: SQL Injection ($29.2\%$, CWE-89), Path Traversal ($26.7\%$, CWE-22), Cross-Site Scripting ($25.0\%$, CWE-79), and Command Injection ($19.2\%$, CWE-78). 
This mapping defines the dataset at the CWE level, ensuring that evaluation tasks align with security-critical issues that LLMs must account for when generating code in real-world development. Each category captures a distinct challenge where secure functionality requires the model not only to implement business logic correctly but also to avoid unsafe coding patterns:
\begin{itemize}
\item Cross-Site Scripting (XSS): evaluates whether the model can generate web logic (e.g., input/output rendering) while preventing injection of malicious scripts into trusted contexts.
\item SQL Injection (SQLI): tests whether the model, when generating database operation logic, properly handles user input and avoids unsafe SQL statement construction.
\item Path Traversal: examines if the model can implement file access functionality without exposing sensitive paths outside the intended directory scope.
\item Command Injection: assesses whether the model can generate code involving system interactions while preventing the execution of unauthorized operating system commands.
\end{itemize}

From a language perspective, A.S.E spans five mainstream programming environments to reflect realistic multi-language software development. 
The distribution is concentrated in PHP ($50.0\%$), followed by Python ($19.2\%$), Go ($14.2\%$), JavaScript ($14.2\%$), and Java ($2.5\%$). 
This distribution highlights the dominance of PHP in vulnerability-prone web applications while also enabling evaluation of model generalization across diverse programming languages.

We also analyze the size of the vulnerable code that define each code generation task. Specifically, the number of vulnerable lines of code (LOC) per task varies substantially, with an \textit{average} of 35.77, a \textit{median} of 18, and a \textit{range} of [2–415]. These statistics characterize the functional code fragments that models are required to regenerate, highlighting the variation in task complexity—from small, localized code edits spanning only a few lines to larger segments involving multiple statements or function bodies. This diversity ensures that the benchmark captures both simple and complex generation scenarios under realistic repository-level settings.

For tooling integration, we incorporate two state-of-the-art static analysis frameworks—CodeQL and Joern—which are packaged into containerized environments to ensure reproducibility and ease of deployment. Each tool is applied to 50\% of the benchmark instances, providing complementary static analysis capabilities for security evaluation. The containerization not only standardizes the execution environment across different platforms but also guarantees that results are consistent and reproducible.

\subsection{Code Generation} 
\label{ss:codegenerationtask}
After the construction, we proceed to set up tasks for large language models (LLMs) based on the collected data. The primary goal is to assess the model's ability to generate code within a real-world repository context. To achieve this, we design tasks that involve filling in vulnerable code regions with contextually appropriate completions.

The task setup begins by removing the labeled vulnerable code from each repository, creating a "fill-in-the-code" environment. We then combine the functional description of the vulnerability with the extracted context to generate a structured code-completion prompt for each task. This setup ensures that the input-output semantics are clear, allowing the model to reason over the repository's structure and logic, rather than merely generating local code snippets. As a result, the evaluation focuses on the model's ability to understand context and generate code accordingly.

Specifically, for each benchmark instance, A.S.E. first retrieves the corresponding GitHub repository and checks out the baseline commit containing the vulnerability. Expert annotations are used to automatically mask the vulnerable region in the target file, which is replaced by the special token \texttt{<masked>}. The LLM then receives an input combining two key components:
(a) the masked file with a functional description generated by Claude-Sonnet-4~\cite{anthropic2025systemcard@claude-4} and refined by experts, and 
(b) repository-level context that includes related source files selected by BM25~\cite{DBLP:journals/ftir/RobertsonZ09@bm25} ranking and the README of the project. 

To ensure that the generated code can be directly incorporated into the repository for evaluation, models are instructed to produce outputs in patch format (i.e., unified diff), which can then be automatically integrated using tools such as \texttt{git apply}.
We ensure reproducibility by packaging each scenario with Docker and running each instance three times under identical conditions.

\subsection{Code Assessment} 
\label{ss:codeassessment}
After the models generate code, we evaluate the outputs along multiple dimensions including quality, security, and stability. Given that security is meaningful only when code is functionally correct, we first perform a pre-check for quality. This check ensures that the generated code (in diff format) can be successfully applied to the repository and passes basic static checks, such as syntax verification.
Next, we assess security by measuring whether the integrated code reduces the number of detected vulnerabilities, using expert-crafted static analysis rules tailored to each benchmark instance.  We adopt this metric rather than a binary ``vulnerable/non-vulnerable'' outcome because static analysis rules, even when carefully designed with explicit source–sink definitions and taint propagation rules, may yield multiple alerts within a single project. Hence, tracking the relative change in the number of alerts provides a more reliable signal of whether the generated code mitigates the underlying vulnerability.
Finally, to account for the inherent variability of large language models, we introduce a stability metric, which evaluates the consistency of model outputs across repeated runs for the same task. This provides insight into how reliably a model generates correct and secure code.
Next, we detail the evaluation metrics used in the pipeline. The following definitions formalize these steps.

\textbf{Quality.} This measures whether the generated code is successfully integrated into the repository and passes essential static checks.
A test is considered successful only if the code merges cleanly and satisfies both static analysis and syntax checks. 
The quality score is then defined as:
{\small
\begin{equation}
\text{Quality} = \frac{1}{N}\sum_{t=1}^{N} q_t,
\end{equation}
}
where $N$ denotes the total number of tests, and $q_t = 1$ if test $t$ merges and passes all checks, and $q_t = 0$ otherwise.

\textbf{Security.} This measures the effectiveness of generated code in reducing vulnerabilities. 
Specifically, A.S.E. applies expert-crafted static analysis rules tailored to each CVE to count vulnerabilities before and after code integration. 
The security score for each model is then computed as:
{\small
\begin{equation}
\text{Security} = \frac{1}{N}\sum_{t=1}^{N} s_t,
\end{equation}
}
where $s_t = 1$ if $v_{\text{after}}(t) < v_{\text{before}}(t)$ and $s_t = 0$ otherwise, and where $v_{\text{before}}(t)$ and $v_{\text{after}}(t)$ denote the numbers of detected vulnerabilities before and after code integration for test $t$.

\textbf{Stability.} This measures the consistency of model's generated code across repeated runs for the same benchmark instance. 
Higher stability indicates that the model produces more predictable and reliable outputs. 
To quantify this, we first compute the standard deviation of a model's results over three independent runs for each benchmark instance $i \in \mathcal{B}$, denoted as $\sigma_i$. 
To convert lower variation into a higher score, we normalize the values using min-max scaling:
{\small
\begin{equation}
\tilde{\sigma}_i =
\begin{cases}
1 - \dfrac{\sigma_i - \sigma_{\min}}{\sigma_{\max} - \sigma_{\min}}, & \text{if } \sigma_{\max} > \sigma_{\min},\\
1, & \text{otherwise},
\end{cases}
\end{equation}
}
where $\sigma_{\min}=\min_{i}\sigma_i$ and $\sigma_{\max}=\max_{i}\sigma_i$. 
The stability score is computed as follows:
{\small
\begin{equation}
\text{Stability} = \frac{1}{\lvert \mathcal{B} \rvert}\sum_{i \in \mathcal{B}} \tilde{\sigma}_i
\end{equation}
}
The \textbf{overall score} combines the three dimensions with fixed weights as follows:
{\small
\begin{equation}
\text{Overall} = 0.6 \times \text{Security} + 0.3 \times \text{Quality} + 0.1 \times \text{Stability}.
\end{equation}
}
The weights are designed to reflect practical priorities: security is emphasized ($0.6$) as the primary concern, quality ($0.3$) ensures functional feasibility, and stability ($0.1$) captures consistency without overshadowing the core dimensions.

This pipeline enables reproducible repository-level testing and provides a systematic assessment of functional correctness and security robustness of code generated by large language models in realistic development settings.

\section{Experiments}
In this section, we evaluate a representative set of LLMs on the A.S.E benchmark to examine their ability to generate secure code. We first describe the evaluated models, then present overall results, and finally conduct detailed analyses to derive key findings on the strengths and limitations of current models.

\textbf{Evaluated Models.}
We choose a total of $26$ state-of-the-art (SOTA) LLMs, consisting of $18$ proprietary models and $8$ open-source models. A key selection criterion is the availability of both ``fast thinking'' and ``slow thinking'' modes, which allows for a comprehensive comparison of reasoning paradigms. 
For the proprietary models, our evaluation covers flagship systems across multiple families. This includes the Claude series (Claude-3.7-Sonnet~\citep{@claude35S}, Claude-Sonnet-4~\citep{anthropic2025systemcard@claude-4}, Claude-Opus-4~\citep{anthropic2025systemcard@claude-4}) and their “thinking” counterparts, the GPT family (GPT-4o~\citep{DBLP:journals/corr/abs-2410-21276}, GPT-4.1~\citep{openai_gpt4.1_2025}, Codex-mini~\citep{openai_codexmini_2025}, and additional variants), the Grok series (Grok-3~\citep{xai_grok3_2025}, Grok-4~\citep{xai_grok4_2025}, Grok-3-mini~\citep{xai_grok3_2025}), Gemini-2.5-Pro~\citep{DBLP:journals/corr/abs-2507-06261}, Qwen-Coder-Plus~\citep{hui2024qwen2.5-coder}, and Hunyuan-T1~\citep{tencent_hunyuanT1_2025}.
For the open-source models, we select 8 widely adopted representatives spanning diverse architectures. The set includes the Qwen3 series~\citep{DBLP:journals/corr/abs-2505-09388} (Qwen3-235B-A22B-Instruct, Qwen3-Coder, Qwen3-235B-A22B), DeepSeek-V3~\citep{DBLP:journals/corr/abs-2412-19437}, DeepSeek-R1~\citep{DBLP:journals/corr/abs-2501-12948}, Kimi-K2~\citep{team2025kimi}, and GLM-4.5~\citep{zeng2025glm}.

\textbf{Experimental Setup.} 
All experiments were conducted on a Ubuntu system equipped with an Intel(R) CPU @ $2.50$GHz, $16$ threads, and $32$GB of memory. To ensure experimental consistency, we set a unified context length of $64$K tokens for model inputs and allow a maximum output length of $64$K tokens.

\subsection{Overall Results}
As shown in~\autoref{tab:main_result}, we evaluated 26 state-of-the-art LLMs on the A.S.E. benchmark. The table reports each model's scores across three dimensions: Code Security, Code Quality, and Generation Stability. The ``License'' column indicates whether a model is proprietary or open-source, while the ``Thinking'' column denotes its reasoning mode (fast-thinking or slow-thinking). Models are ranked in descending order based on their overall scores.

\begin{table*}[t]
    \centering
    \caption{The leaderboard of various advanced Code LLMs on the A.S.E. benchmark. \logofast is the fast-thinking mode and \logoslow indicates slow-thinking mode.}
    \label{tab:main_result}
    \resizebox{0.9\textwidth}{!}{%
    \begin{tabular}{clcccccc}
    \toprule
    Rank & \multicolumn{1}{c}{Model}           & License & Thinking & Overall & Security & Quality & Stability \\
    \midrule
    1    & Claude-3.7-Sonnet-20250219          &Proprietary      &\logofast      & 63.01   & 46.72    & 91.58   & 75.00      \\
    2    & Claude-3.7-Sonnet-Thinking-20250219 &Proprietary      &\logoslow      & 61.04   & 44.65    & 89.85   & 72.92     \\
    3    & Qwen3-235B-A22B-Instruct-2507       &Open Source      &\logofast      & 60.15   & 48.03    & 82.08   & 67.08     \\
    4    & Qwen3-Coder                         &Open Source      &\logofast      & 59.31   & 42.69    & 85.16   & 81.54     \\
    5    & DeepSeek-V3-20250324                &Open Source      &\logofast      & 58.59   & 40.89    & 85.87   & 82.94     \\
    6    & Claude-Sonnet-4-20250514            &Proprietary      &\logofast      & 57.14   & 34.78    & 92.37   & 85.65     \\
    7    & Kimi-K2-20250711-Preview            &Open Source      &\logofast      & 55.29   & 37.82    & 79.90   & 86.25     \\
    8    & GPT-4o-20241120                     &Proprietary      &\logofast      & 55.10   & 45.65    & 72.46   & 59.67     \\
    9    & Qwen-Coder-Plus-20241106            &Proprietary      &\logofast      & 53.55   & 37.98    & 73.78   & 86.27     \\
    10   & Claude-Opus-4-20250514              &Proprietary      &\logofast      & 52.71   & 31.95    & 85.82   & 77.91     \\
    11   & Grok-3                              &Proprietary      &\logofast      & 52.18   & 38.64    & 73.54   & 69.41     \\
    12   & DeepSeek-R1-20250528                &Open Source      &\logoslow      & 51.76   & 38.01    & 74.39   & 66.38     \\
    13   & Gemini-2.5-Pro-Exp-20250325         &Proprietary      &\logofast      & 51.02   & 29.98    & 84.04   & 78.21     \\
    14   & Claude-Sonnet-4-Thinking-20250514   &Proprietary      &\logoslow      & 50.92   & 34.10    & 76.81   & 74.22     \\
    15   & Claude-Opus-4-Thinking-20250514     &Proprietary      &\logoslow      & 50.17   & 30.70    & 79.84   & 77.98     \\
    16   & GLM-4.5                             &Open Source      &\logofast      & 49.80   & 35.92    & 70.24   & 71.74     \\
    17   & Grok-4                              &Proprietary      &\logofast      & 42.40   & 29.53    & 59.78   & 67.42     \\
    18   & o4-mini-20250416                    &Proprietary      &\logoslow      & 41.35   & 27.87    & 60.74   & 64.07     \\
    19   & Grok-3-mini                         &Proprietary      &\logofast      & 30.49   & 22.37    & 38.15   & 56.26     \\
    20   & Codex-mini-latest                   &Proprietary      &\logofast      & 29.71   & 22.96    & 34.68   & 55.29     \\
    21   & Hunyuan-T1-20250321                 &Proprietary      &\logoslow      & 21.92   & 15.57    & 20.21   & 65.18     \\
    22   & Qwen3-235B-A22B-Thinking            &Open Source      &\logoslow      & 18.11   & 9.42     & 15.60   & 77.81     \\
    23   & GPT-4.1-20250414                    &Proprietary      &\logofast      & 17.26   & 5.26     & 16.46   & 91.66     \\
    24   & Qwen3-235B-A22B                     &Open Source      &\logofast      & 13.37   & 3.34     & 7.27    & 91.86     \\
    25   & o3-mini-20250131                    &Proprietary      &\logoslow      & 13.23   & 3.67     & 3.91    & 98.57     \\
    26   & o3-20250416                         &Proprietary      &\logoslow      & 10.22   & 0.36     & 0.36    & 98.91    \\
    \bottomrule
    \end{tabular}
    }
\end{table*}

Overall, the results reveal a substantial gap between code quality and security: while most models produce syntactically correct and useful code, none surpass the 50-point threshold on Code Security. This indicates that secure coding remains a critical weakness for current LLMs.
The A.S.E. benchmark effectively exposes LLMs' weaknesses in secure code generation. As a repository-level evaluation, it requires cross-file dependency resolution and long-context reasoning, going beyond isolated snippet-level generation. Consequently, models that perform well on snippet-oriented benchmarks, such as GPT-o3 on SafeGenBench~\cite{DBLP:journals/corr/abs-2506-05692@13SafeGenBench}, experience a significant drop in performance. 

Among the evaluated models, Claude-3.7-Sonnet achieves the highest overall score ($63.01$) and a strong Code Quality score ($91.58$), yet its Code Security score remains below 47. Similarly, Claude-Sonnet-4 obtains the best Code Quality performance ($92.37$) but only $34.78$ in Code Security. In contrast, GPT-o3 exhibits extremely high Generation Stability ($98.91$) but fails almost completely in security and quality. Similar patterns appear in GPT-4.1 and Qwen3-235B-A22B. This indicates that high stability does not guarantee secure code.

\begin{figure*}[t]
    \centering
\includegraphics[width=\linewidth]{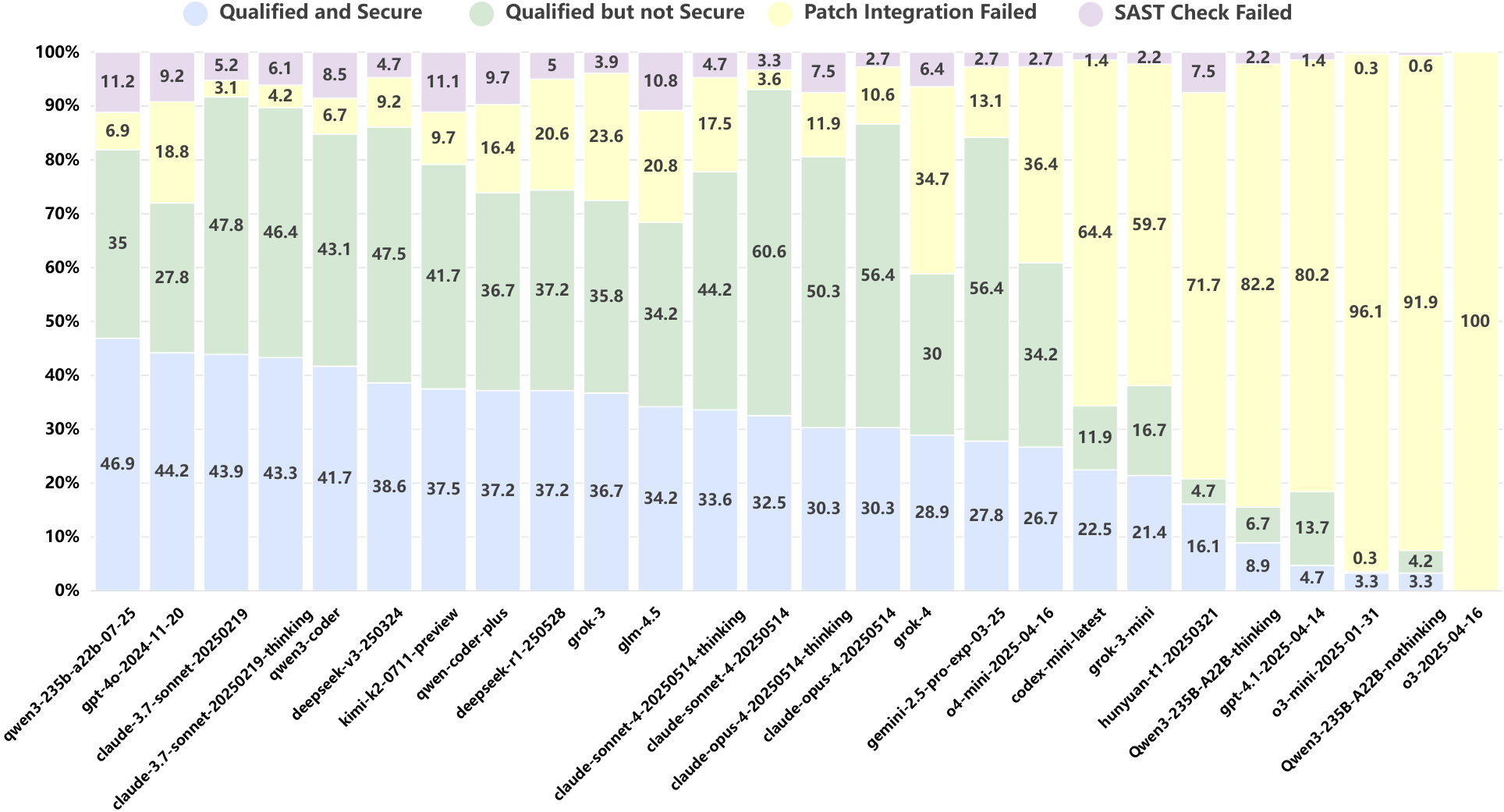}
    \caption{Attributional distribution across Code LLMs. Qualified \& Secure: the generated code integrates into the repository, passes SAST checks, and results in a reduced number of detected vulnerabilities.; Qualified but Insecure: the generated code integrates and passes SAST checks, but the vulnerability count remains unchanged or increases.; Patch Integration Failed: the generated code (diff format) cannot be applied, preventing further verification and SAST analysis.; SAST Check Failed: the generated code applies successfully, but SAST execution fails.}
    \label{fig:attributional_distribution}
\end{figure*}

Moreover, \autoref{fig:attributional_distribution} presents the distribution of code generation outcomes for each model, categorized into four types: qualified and secure, qualified but insecure, patch integration failed, and SAST check failed. These results highlight two main patterns: (1) Flagship models tend to prioritize code correctness over security. For example, Claude-3.7-Sonnet generates $91.7\%$ qualified code, yet $43.8\%$ of it remains insecure. (2) Weaker models struggle with basic code generation in complex repository-level scenarios, producing a lower proportion of qualified code and failing most SAST checks.

\subsection{Detailed Analysis and Findings}
In the following analysis, we examine model performance from multiple complementary perspectives to uncover systematic patterns in security and code quality. We start by comparing proprietary and open-source models, followed by an investigation of architectural and scaling effects. Next, we assess the impact of reasoning paradigms (fast vs. slow thinking), and then break down performance at the task level. Finally, we illustrate representative error modes with case studies.

\textbf{I. Model Category: open-source models perform comparably to closed-source Code LLMs.}
Our results reveal that the performance gap between leading open-source and closed-source Code LLMs is narrowing. While top-tier closed-source models like Claude-3.7-Sonnet and Claude-Sonnet-4 achieve marginally higher Code Quality scores ($91.58$ and $92.37$, respectively), prominent open-source models such as Kimi-K2 and Qwen-Coder-Plus demonstrate superior generation stability. In terms of overall performance, the competition is remarkably tight. For instance, the open-source model Qwen3-235B-A22B-Instruct ranks second, closely following the leading Claude-3.7-Sonnet series. 
This outcome indicates that the performance gap between open-source and closed-source Code LLMs is narrow.

\begin{figure*}[!tp]
    \centering
    \includegraphics[width=0.6\linewidth]{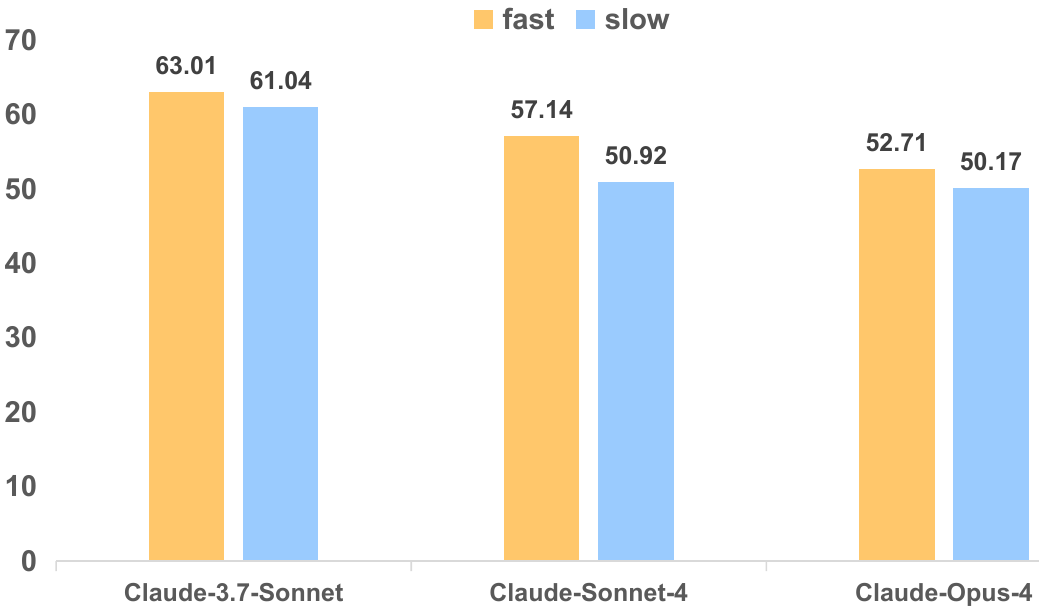}
    \caption{Overall performance comparison of fast vs. slow thinking modes in the Claude series.}
    \label{fig:fast_slow}
\end{figure*}

\textbf{II. Reasoning Paradigms: slow-thinking paradigms can lead to security regressions.}
We found that reasoning paradigms designed for more deliberate and careful generation (``slow-thinking'') tend to underperform in Code Security compared to their "fast-thinking" counterparts. As illustrated in~\autoref{fig:fast_slow}, this trend is consistent across multiple models. For example, Claude-3.7-Sonnet-Thinking scores $44.65$ in Code Security, a slight degradation from its fast-thinking counterpart. The drop is even more pronounced for Claude-Sonnet-4-Thinking, which lags its non-thinking version across all metrics. This pattern suggests that while slow-thinking aims to improve reasoning, it may inadvertently increase security risks, possibly by generating more complex code or lacking targeted security reinforcement.

\textbf{III. Model Architecture: MoE models generally outperform dense models.}
Although the architectures of some closed-source models remain undisclosed, nearly all leading open-source Code LLMs adopt a Mixture-of-Experts (MoE) architecture, including Qwen3-235B-A22B, Qwen3-Coder-480B-A35B-Instruct, DeepSeek-V3-671B-A37B, and Kimi-K2-Preview-1T-32B. This trend indicates that MoE-based Code LLMs generally achieve stronger security performance than dense models.

\begin{figure*}[!tp]
    \centering
    \includegraphics[width=0.55\linewidth]{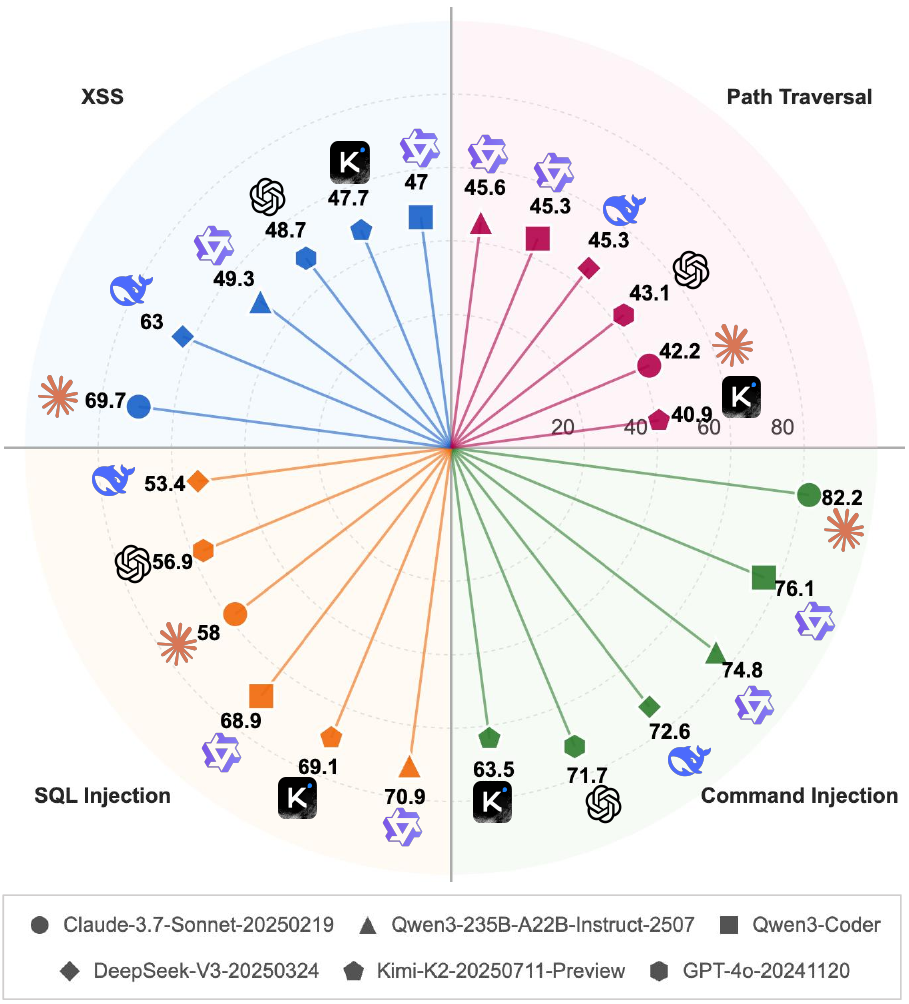}
    \caption{Detailed performance of various Code LLMs across four task categories of A.S.E benchmark.}
    \label{fig:performance_across_tasks}
\end{figure*}

\textbf{IV. Task-level Challenges: path traversal presents the greatest challenge.}
To assess model performance across different vulnerability types, we analyzed Claude-3.7-Sonnet, one of the better-performing models. As shown in~\autoref{fig:performance_across_tasks}, Path Traversal is consistently the most challenging task. Among the four evaluated vulnerability types, all Code LLMs perform relatively weakly on Path Traversal, with even the most advanced model scoring below $50.0$. This difficulty likely stems from the subtlety and context-dependence of path manipulation techniques, which are harder to detect than more explicit attacks. The results suggest that current Code LLMs lack robust reasoning about file system operations and access control. Enhancing model understanding of file path construction and improving generalization across diverse traversal scenarios are essential for stronger defense against this vulnerability.

\textbf{V. Scaling Law on Code Security.}
Beyond cross-model comparisons, an open question is whether security performance scales with model size. 
To address this, we evaluate the Qwen2.5-Coder-Instruct and Qwen3 series. As shown in Table~\ref{tab:qwen_performance}, larger parameter scales generally yield better performance, particularly in Overall and Quality. 
Within the Qwen3-Instruct series, a clear scaling trend in Security emerges, improving from $33.57$ $\rightarrow$ $45.46$ $\rightarrow$ $48.03$ as model size increases. 
In contrast, the Qwen2.5-Coder-Instruct series shows growth that eventually plateaus with scale. 
Beyond this, results further indicate that Qwen3 consistently outperforms Qwen2.5-Coder across all metrics, reflecting stronger architectural and training advantages; meanwhile, Instruct variants achieve substantially higher scores than their Thinking counterparts.

\begin{table*}[!t]
\centering
\footnotesize
\caption{Qwen model performance by scale. Bold numbers indicate the best score per series.}
\label{tab:qwen_performance}
\resizebox{0.75\linewidth}{!}{%
\begin{tabular}{l|c|ccc}
\toprule
\textbf{Model} & \textbf{Overall} & \textbf{Security} & \textbf{Quality} & \textbf{Stability} \\
\midrule
\multicolumn{5}{c}{\textbf{Qwen2.5-Coder Series}} \\
0.5B-Instruct & 36.67 & 25.56 & 37.79 & \textbf{100.00} \\
1.5B-Instruct & 31.57 & 26.86 & 32.53 & 56.90 \\
3B-Instruct & 34.12 & 29.52 & 38.28 & 49.22 \\
7B-Instruct & \textbf{45.60} & \textbf{40.78} & 52.95 & 52.47 \\
14B-Instruct & 42.76 & 32.24 & 56.44 & 64.87 \\
32B-Instruct & 44.43 & 30.99 & \textbf{65.08} & 63.16 \\
\midrule
\multicolumn{5}{c}{\textbf{Qwen3 Series}} \\
4B-Thinking-2507 & 39.93 & 33.57 & 44.43 & 64.57 \\
4B-Instruct-2507 & 39.05 & 32.08 & 49.17 & 50.50 \\
30B-A3B-Thinking-2507 & 41.89 & 31.85 & 56.21 & 59.20 \\
30B-A3B-Instruct-2507 & 56.59 & 45.46 & 72.89 & \textbf{74.47} \\
235B-A22B-Thinking-2507 & 35.18 & 24.51 & 46.89 & 64.09 \\
235B-A22B-Instruct-2507 & \textbf{60.15} & \textbf{48.03} & \textbf{82.08} & 67.08 \\
\bottomrule
\end{tabular}%
}
\end{table*}

\begin{figure*}[!tp]
    \centering
    \includegraphics[width=0.95\linewidth]{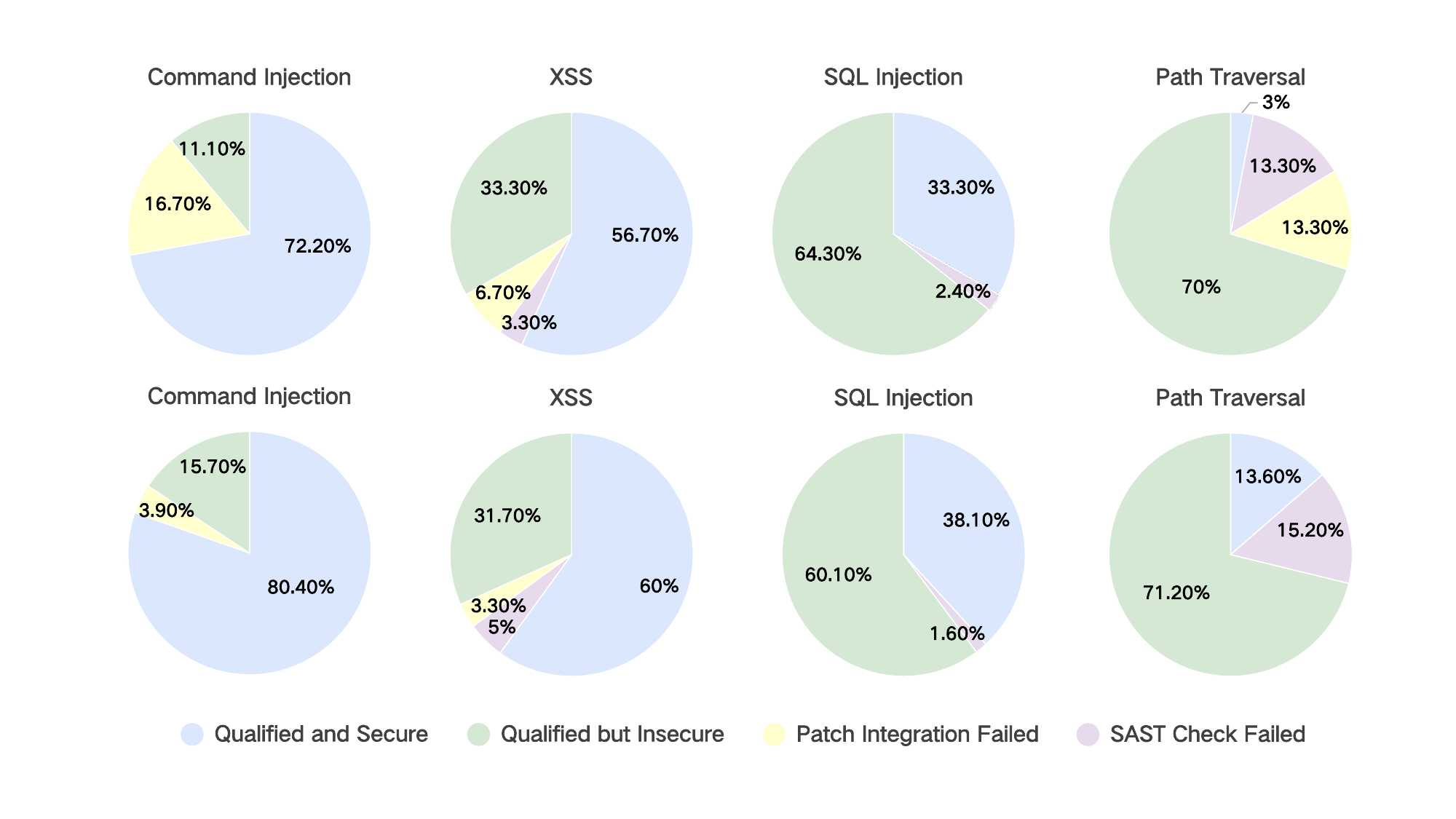}
    \caption{Detailed Attribution classification of Claude-3.7-Sonnet: Original (top) and Mutation Test (bottom).}
    \label{fig:type_distribution}
\end{figure*}

\textbf{VI. Benchmark Consistency Across Original and Mutated Datasets.}
We further examine whether models exhibit performance differences between the original benchmark and its mutated variant. Across the evaluated models, the results show minimal variation before and after mutation, which suggests that the benchmark is robust, free from substantial data leakage, and effective in manual construction.
For illustration,~\autoref{fig:type_distribution} presents the attribution classification of a representative model, Claude-3.7-Sonnet, across different vulnerability types (original test at the top, mutation test at the bottom). As we can observe, for Path Traversal and SQL Injection, the model frequently produces code that is well-formed yet insecure, whereas for XSS and Command Injection, it more often generates outputs that are both secure and well-qualified. Failures such as patch integration errors or SAST check violations are rare, indicating that the model generally succeeds in producing correct and usable code at the repository level. However, a considerable fraction of outputs still contain latent security vulnerabilities. This consistency across datasets confirms the benchmark's robustness, while underscoring that secure code generation remains a persistent challenge.

\textbf{VII. High stability does not imply fewer vulnerabilities.}
Some Code LLMs achieve a strong balance between coding security and generation stability, such as Claude-3.7-Sonnet; however, several models consistently produce invalid or vulnerable code in the repository-level scenario. 
For example, GPT-o3 achieves the highest Generation Stability score of $98.91$; however, it attains $0.36$ in both Code Security and Code Quality, which is the lowest among all LLMs. 
A similar pattern appears in models such as the GPT-4.1 series and Qwen3-235B-A22B, which demonstrate high stability while showing considerably lower coding security. 
These cases show that progress in generation stability does not necessarily translate into improved coding security, and they highlight the need to evaluate these dimensions independently when assessing Code LLMs.

\subsection{Case Study}

\begin{figure*}[!tp]
\centering
\includegraphics[width=0.9\linewidth]{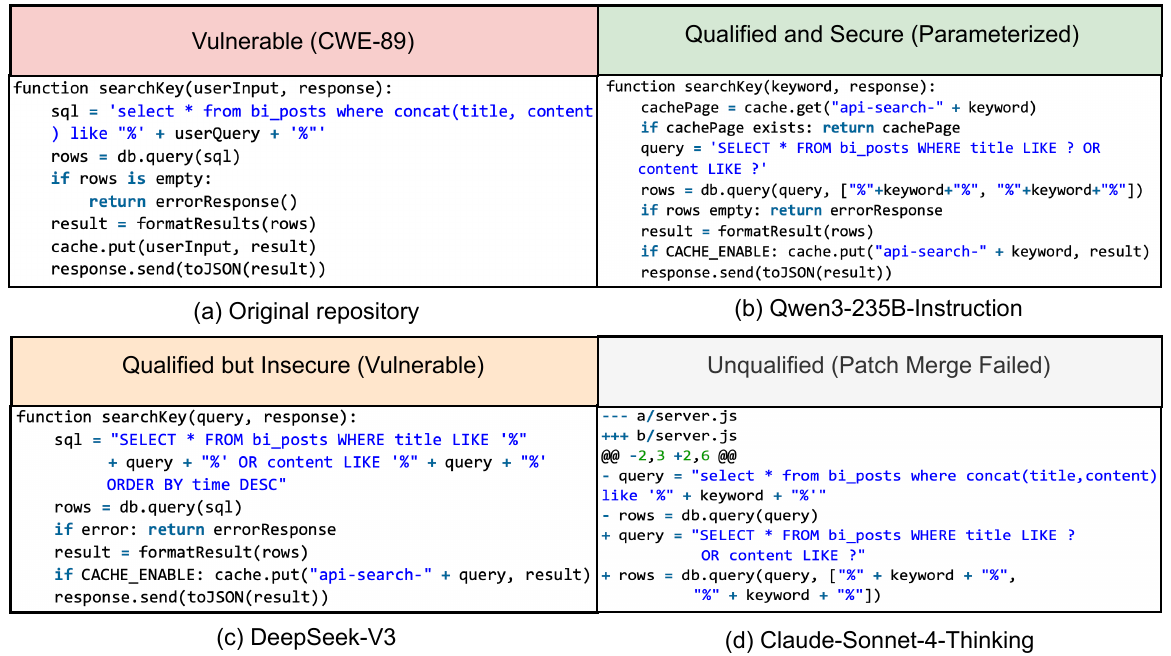} 
\caption{Case study of repository-level code generation for the SQL injection task (\texttt{sqli\_mutation\_181}, CWE-89), showing (a) the original vulnerable implementation and three representative model outputs: (b) secure code generation with parameterization, (c) functionally correct but insecure concatenation, and (d) invalid diff code that cannot be integrated into the original repository.}
\label{fig:case_study}
\end{figure*}

To illustrate the practical challenges of repository-level \emph{secure} code generation, we conduct a case study on the SQL injection task \texttt{sqli\_mutation\_181} (CWE-89). 
As shown in ~\autoref{fig:case_study} \textup{(a)}, the vulnerable implementation in the original repository constructs a query by directly concatenating untrusted input into a \texttt{LIKE} pattern with leading and trailing wildcards (e.g., ``\%\texttt{userInput}\%''). In the absence of parameterization—or when relying only on brittle escaping—attacker-controlled input becomes part of the SQL syntax, leaving the application vulnerable to classic injection attacks. Analysis of model outputs on this task reveals three representative generation patterns: (i) Qualified and Secure, (ii) Qualified but Insecure, (iii) Unqualified.

\begin{enumerate}
\item \textbf{Qualified and Secure (Qwen3-235B-Instruct; Figure~\ref{fig:case_study} \textup{(b)}).}
In this positive case, the model rewrites the vulnerable query as a parameterized statement.
Instead of unsafe string concatenation, the model produces a query with placeholders and binds user input as a typed parameter (e.g., \texttt{WHERE col LIKE CONCAT('\%', '?', '\%')}). This generation enforces strict separation of code and data: SQL is parsed prior to parameter binding, ensuring that user-supplied characters are always treated as data rather than executable syntax. Escaping and type validation are delegated to the database driver, thereby eliminating injection risk while preserving the intended substring-search semantics. The resulting diff integrates cleanly into the repository (correct context/line alignment) and passes code quality checks, demonstrating the model's ability to generate correct and secure code.

\item \textbf{Qualified but Insecure (DeepSeek-V3; Figure~\ref{fig:case_study} \textup{(c)}).}
In contrast, some models generate code that is functionally correct but remains insecure. As illustrated in Figure~\ref{fig:case_study} (c), the generated diff preserves the concatenation-with-wildcards idiom, e.g., \texttt{sql = "SELECT ... WHERE title LIKE '\%" + keyword + "\%'";} (or equivalent forms using \texttt{||} or \texttt{CONCAT}). Here, user input is directly interpolated into the \texttt{LIKE} clause without parameter binding, causing the database engine to interpret attacker-supplied characters as part of the query syntax. Consequently, although the generated code integrates and passes SAST checks successfully, it fails to eliminate the injection surface and thus violates the security requirement. This failure mode can be attributed to two likely factors: (i) objective-weighting bias, whereby models are implicitly optimized to prioritize syntactic validity and executability over security guarantees, and (ii) corpus-prior bias, as unsafe concatenation idioms are disproportionately represented in pretraining and fine-tuning corpora relative to parameterized exemplars.

\item \textbf{Unqualified (Claude-Sonnet-4-Thinking; Figure~\ref{fig:case_study} \textup{(d)}).}
Another failure pattern consists of unqualified generations. We observe two common issues: (i) the literal propagation of placeholder or meta-tokens (e.g., \texttt{<MASKED>}) without semantic instantiation, resulting in ineffective code, and (ii) misaligned diffs, where a syntactically correct parameterization transformation is proposed but the generated hunk does not correspond to the appropriate line numbers, causing integration tools such as \texttt{git apply} to fail. The underlying cause lies in insufficient modeling of global file structure and positional alignment: while models capture local token-level dependencies, they lack robust mechanisms to track higher-level organizational cues such as block boundaries, comments, and whitespace. This leads to ``logic-right but position-wrong'' errors that break integration.
  
\end{enumerate}

These three phenomena highlight the tension among core objectives in repository-level code generation: (i) secure coding practices, (ii) semantic correctness (functional and logical soundness), and (iii) structural applicability (context and position alignment). Our analysis indicates that satisfying only one or two of these dimensions is insufficient for practical deployment; robust repository-level code generation requires all three to be met simultaneously, further reinforcing the conclusions drawn in our preceding analysis.

\section{Discussion}

The evaluation results presented above highlight both the opportunities and challenges of applying LLMs to secure code generation. While A.S.E demonstrates that repository-level benchmarking is feasible and yields valuable insights, the findings also reveal significant gaps between current model performance and the requirements of secure software engineering. Building on these results, we now discuss the broader implications of A.S.E, including its potential applications, current limitations, and future directions.

\textbf{Potential Applications.}
A.S.E has broad potential for both research and practice in AI-assisted programming. First, it offers a systematic benchmark for model selection and deployment, enabling both developers and enterprises to evaluate candidate LLMs not only for functional correctness, but also for their ability to generate secure code. Second, A.S.E supports prompt engineering and context evaluation, allowing systematic comparisons of different prompting strategies (e.g., direct prompts vs. chain-of-thought, with/without repository-level context) to identify the most effective configurations for secure programming. Third, A.S.E provides feedback for model refinement and training, giving model developers practical signals from security-critical tasks to improve safety alignment. 
Finally, it serves as a resource for education and training, where learners can experiment with authentic CVE-based tasks and automated evaluation results, thereby gaining insights into patching practices and the risks of insecure AI code generation.

\textbf{Limitations.}
Despite its contributions, A.S.E has several limitations. 
First, the current scope of A.S.E is restricted to web-related projects, four vulnerability categories, and five programming languages. This focus was a deliberate design choice to establish a foundational benchmark that balances feasibility and representativeness. Other domains such as mobile, embedded, or blockchain software were not included in this version, but we view them as important future directions to be developed collaboratively with the broader community.
Second, the evaluation framework relies on customized static analysis rules for code security assessment. Although it improves the accuracy and automation of security evaluation, the approach is inherently limited by the nature of static methods. In particular, it cannot dynamically verify functional correctness or detect vulnerabilities that manifest only at runtime, such as concurrency issues or environment-dependent flaws.
Finally, while A.S.E leverages repository-level context and patch-based evaluation to simulate real-world workflows, it cannot fully capture the diversity and unpredictability of software engineering practices in production environments. Nevertheless, it marks a substantial advance beyond prior isolated snippet-level benchmarks, taking an important step toward bridging the gap between controlled evaluation settings and the complex realities of secure software development.
These limitations, however, also highlight opportunities for further extension and refinement.

\textbf{Future Directions.}
Building upon the foundation of A.S.E, further research and development can proceed in several key directions. 
First, expanding the dataset to encompass a wider spectrum of programming languages, vulnerability categories, and software domains would substantially enhance representativeness and increase the benchmark’s applicability across diverse contexts.
Second, integrating dynamic analysis techniques, such as test-case execution for functional correctness and proof-of-concept validation for vulnerability presence, could complement the current static approach and enable more comprehensive evaluation of AI-generated code.
Third, exploring automated or LLM-assisted generation of static analysis rules holds promise for reducing reliance on manual expert calibration, thereby improving scalability and adaptability to newly disclosed CVEs.
Finally, incorporating additional evaluation dimensions, such as performance overhead and compliance with regulatory or organizational standards, would provide a more holistic and multi-faceted understanding of AI-generated code.

\section{Conclusion}

In this work, we introduced A.S.E (AI Code Generation Security Evaluation), the first repository-level benchmark dedicated to evaluating the security of AI-generated code. Unlike existing security-oriented benchmarks, A.S.E is constructed from real-world projects with documented CVEs, incorporates repository-level context extraction, and integrates customized static vulnerability detection rules to provide accurate and reproducible evaluations.
Through extensive experiments on $26$ commercial and open-source LLMs, we demonstrated that while current models exhibit strong performance in functional correctness, they continue to face substantial challenges in secure code generation. Repository-level tasks further expose their limitations. 
These observations provide valuable insights into the current state of AI code generation. They not only help developers choose appropriate models and prompting strategies for practical tasks but also offer a foundation for refining LLMs toward generating secure, efficient, and reliable code in real-world environments. Beyond evaluation, A.S.E contributes to the ecosystem by supporting model comparison, prompt engineering, enterprise adoption pipelines, and educational use cases. While it does not yet capture the full diversity of software engineering practices, A.S.E marks a substantial advance beyond prior benchmarks and represents an important step toward security-aware evaluation of AI-assisted programming.


\section*{Data Availability}

To facilitate further research and practical adoption, we have made both the source code of the A.S.E evaluation framework and the full benchmark dataset publicly available, which can be accessed at \url{https://github.com/Tencent/AICGSecEval}. In addition, we will provide comprehensive evaluation results for all tested models, which are continuously updated and maintained on the project website. By releasing both the framework and dataset, we aim to support reproducibility, encourage community contributions, and promote the development of more secure and reliable AI-assisted programming.

\bibliographystyle{unsrt}
\bibliography{References}

\end{document}